\documentclass{aa}

\usepackage[varg]{txfonts}
\usepackage{siunitx}
\usepackage[version=4]{mhchem}
\usepackage{multirow}
\usepackage{graphicx}
\usepackage[caption=false]{subfig}
\usepackage[strict]{changepage}
\usepackage{xcolor}
\usepackage{ulem}
\usepackage[colorlinks=true,citecolor=blue]{hyperref}
\usepackage{float}
\usepackage{amssymb}
\usepackage{amsmath}
\usepackage{lscape}
\usepackage{mathtools}
\usepackage{natbib}
\usepackage{pdflscape}
\usepackage{gensymb}

\defcitealias{mazumdar2021}{Paper~I}   

\DeclareSIUnit\jansky{Jy}

\newcommand{\mum}{$\mu$m }
\newcommand{\coa}{$^{12}$CO }
\newcommand{\cob}{$^{13}$CO }
\newcommand{\coaj}{$^{12}$CO ($3\rm{-}2$) }
\newcommand{\cobj}{$^{13}$CO ($3\rm{-}2$) }

\newcolumntype{L}[1]{>{\raggedright\let\newline\\\arraybackslash\hspace{0pt}}m{#1}}
\newcolumntype{C}[1]{>{\centering\let\newline\\\arraybackslash\hspace{0pt}}m{#1}}
\newcolumntype{R}[1]{>{\raggedleft\let\newline\\\arraybackslash\hspace{0pt}}m{#1}}

\title {High resolution LAsMA $^{12}$CO and $^{13}$CO observation of the G305 giant molecular cloud complex : II. Effect of feedback on clump properties}

\author{P. Mazumdar\inst{\ref{inst1}}
\and F. Wyrowski \inst{\ref{inst1}} 
\and J.\,S.\,Urquhart \inst{\ref{inst2}}
\and D. Colombo \inst{\ref{inst1}}
\and K. M. Menten\inst{\ref{inst1}}
\and S. Neupane \inst{\ref{inst1}}
\and M.\,A.\,Thompson \inst{\ref{inst3}}
}

\institute{Max-Planck-Institut f\"{u}r Radioastronomie, Auf dem H\"{u}gel 69, 53121 Bonn, Germany \label{inst1} \\
\email{pmazumdar@mpifr-bonn.mpg.de}
\and
Centre for Astrophysics and Planetary Science, University of Kent, Canterbury, CT2 7NH, UK \label{inst2}
\and
Centre for Astrophysics Research, Science and Technology Research Institute, University of Hertfordshire, College Lane, Hatfield AL10 9AB, UK \label{inst3}
}

\date{Received xxx / Accepted xxx}
\abstract
{Understanding the effect of feedback from young massive stars on the star forming ability of their parental molecular clouds is of central importance for studies of the interstellar medium and star formation.}
{We observed the G305 star-forming complex in the $J=3\text{-}2$ lines of \ce{^{12}CO} and \ce{^{13}CO} to investigate whether feedback from the central OB stars were triggering star formation in G305 or actually disrupting this process.}
{The region was decomposed into clumps using dendrogram analysis. A catalog of the clump properties such as their positions, luminosities, masses, radii, velocity dispersions, volume densities, surface mass densities, etc. was created. The surface mass densities of the clumps were plotted as a function of the incident 8\,\mum flux. A mask of the region with 8\,\mum flux $> 100\,$MJy/sr was created and clumps were categorized into three classes based on their extent of overlap with the mask, namely ``mostly inside'' (>67\% overlap), ``partly inside'' (>10\% and <67\% overlap), and ``outside'' (<10\% overlap). The surface mass density distribution of each of these populations was separately plotted. This was followed by comparing the G305 clumps with the Galactic average taken from a distance limited sample of ATLASGAL and CHIMPS clumps. Finally, the cumulative distribution functions (CDF) of the clump masses in G305 and their $L/M$ ratios were compared to that of the Galactic sample to determine which mechanism of feedback was dominant in G305. }
{The surface mass densities of clumps showed a positive correlation with the incident 8\,\mum flux.  The data did not have sufficient velocity resolution to discern the effects of feedback on the linewidths of the clumps. The sub-sample of clumps named ``mostly inside'' had the highest median surface mass densities followed by the ``partly inside'' and ``outside'' sub-samples. The difference between the surface mass density distribution of the three sub-samples were shown to be statistically significant using the KS test. The ``mostly inside'' sample also showed the highest level of fragmentation compared to the other two sub-samples. These prove that the clumps inside the G305 region are triggered. The G305 clump population is also statistically different from the Galactic average population, the latter approximating that of a quiescent population of clumps. This provided further evidence that redistribution was not a likely consequence of feedback on the GMC. The CDFs of clump masses and their $L/M$ ratios are both flatter than that of the Galactic average, indicating that clumps are heavier and more efficient at forming stars in G305 compared to the Galactic average.}
{Feedback in G305 has triggered star formation. The collect and collapse method is the dominant mechanism at play in G305.}


\begin{document}

\titlerunning{G305 Giant Molecular Cloud : II. Clump properties}
\authorrunning{P. Mazumdar et al.}
\maketitle

\section{Introduction \label{sec:intro}}

\begin{figure*}
     \centering
     \includegraphics[width=0.85\textwidth]{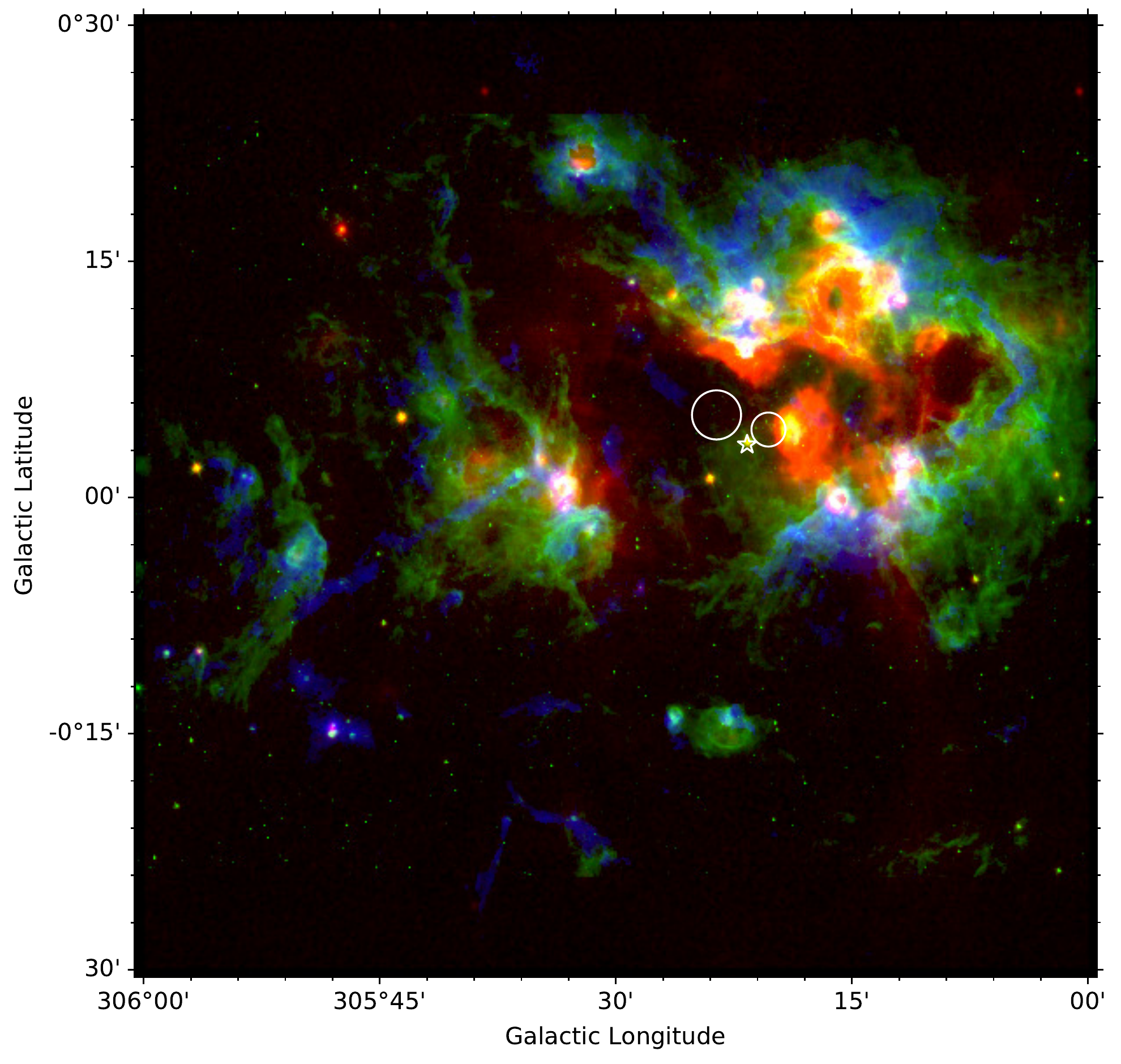}
     \caption{Three-color image (green = \textsl{Spitzer}-IRAC4 8\,$\mu \rm{m}$, red = Midcourse Space Experiment (MSX) 21.3\,$\mu \rm{m}$, blue = LAsMA (\ce{^{13}CO}(3-2) moment-0)) of the G305 star-forming complex. The 21.3\,$\mu \rm{m}$ emission is dominated by hot dust in the HII region. The colder gas is traced by the \ce{^{13}CO}(3-2) line emission. The interface between the ionized and molecular gas appears as a blend of green (strong 8\,$\mu \rm{m}$ emission from PAHs), blue (colder molecular gas) and occasionally red (interfaces very close to HII regions). The positions of Danks 1 and 2 clusters have been marked with the smaller and the larger white circles respectively, and the wolf-rayet star WR48a has been marked as a white star.}
     \label{fig:intro-fig}
    \end{figure*}

    G305 is one of the most massive and luminous star forming giant molecular clouds (GMC) in the Milky Way (\citet{clark}, Fig. \ref{fig:intro-fig}). It is located $\sim \, 4\,\si{kpc}$ \citep{clark, davies12,borissova} away from us at $l\sim\ang{305}\,,\, b\sim\ang{0}$ in the Scutum-Crux spiral arm of the Galactic plane. Its center has been cleared of the interstellar molecular gas by the stellar winds and the ionizing front originating from massive stars belonging to two visible central clusters (Danks 1 and 2) consisting of 21 OB stars and the Wolf-Rayet star (WR48a) \citep{clark, davies12}. A diffuse population of evolved massive stars has also been found to exist inside the cavity with ages similar to that of the two clusters \citep{leistra2005,shara, mauerhan,davies12,faimali,borissova}. The central cavity is also filled with ionized gas as shown by radio continuum observations by \citet{hindson12}. Surrounding the central cavity is a thick layer of molecular gas \citep{hindson10,hindson13,mazumdar2021}. Star formation tracers (water and methanol masers, HII and ultracompact (UC) HII regions and massive young stellar objects (MYSOs)) have also been reported by multiple studies \citep{clark,hindson12,lumsden2013,urquhart2014rms,green09,green_MMB}. Figure \ref{fig:intro-fig} shows a three-color composite image of G305 highlighting three aspects of the complex. The 21.3\,\mum Midcourse Space Experiment (MSX) \citep{price2001} image is shown in red and traces the hot dust being heated by the OB stars within the cavity and towards the edges where the ionization front is interacting with the surrounding molecular gas. The cold molecular gas traced by the integrated \cobj line emission \citep[referred to as Paper I from here on]{mazumdar2021} can be seen in blue being shrouded by emission from the 8\,\mum filter of the Spitzer Infrared array camera (IRAC)  \citep{Churchwell_glimpse} shown in green. The latter traces the polycyclic aromatic hydrocarbons (PAHs) whose emission is excited by the far-UV photons emanating from the stars in the central cavity.
    
    \par The role played by feedback from high mass stars ($M>8\,\rm{M_{\odot}}$) on their surrounding interstellar medium (ISM) has been a subject of active research. As is the case in G305, these stars exist in massive clusters located inside GMCs \citep{motte2018}. They inject energy and momentum into their natal clouds via very strong stellar winds, ionizing radiation and eventually, supernovae \citep{krumholz2014}. Such feedback from OB stars can potentially enhance and/or disrupt the formation of subsequent generations of stars \citep{krumholz2014, deharveng2010}, consequently playing a vital role in the evolution of GMCs \citep{zinnecker2007}. The G305 region, being a very extensively studied GMC, is a very good candidate to examine the effects of feedback on the GMC. Although, circumstantial evidences of triggering in G305 exists \citep{clark, hindson10}, it has, as yet, not been definitively proven.
    
    \par We observed G305 with the Atacama Pathfinder EXperiment (APEX) telescope to examine what effect this feedback from the central population of stars has on the molecular gas content of the GMC \citepalias{mazumdar2021}. The molecular gas in the GMC was mapped in the \coa and \cobj lines. The temperature and column density of the gas facing the center of the cavity was found to be at least a factor of 2 higher than that of the part of the cloud facing away from it, demonstrating that the gas impacted by feedback gets heated and compressed and in turn shields the layers behind it. The rotational excitation of the gas traced by the  $3\rm{-}2$/$2\rm{-}1$ \cob\ line ratios showed similar enhancements at the edge of the cloud facing the center. Using 8\,\mum flux of the region as a proxy to the feedback strength, we demonstrated that regions experiencing feedback had higher median excitation temperatures, rotational excitation, and column density. Examination of the probability distribution function (PDF) of the centroid velocities revealed exponential wings and a stacked spectral analysis also revealed the emission profiles to be positively skewed (which we hypothesized was indicative of the gas being pushed away from us), all of which is consistent with the hypothesis that star formation has been triggered.
    
    \par Having investigated the effects of feedback on the GMC as a whole, we are now interested in exploring the effects of feedback on the ability of the GMC to form subsequent generations of stars. GMCs show a hierarchical structure with clouds condensing into clumps that condense and fragment into cores, eventually forming stars \citep{blitz1986, lada1992}. Since clumps provide the environment and the raw material for star formation, the impact of feedback on the future generation of stars can be studied by their effect on the clump properties in a GMC. In this paper we will decompose the molecular emission distribution in G305 into clumps and investigate their properties in an effort to study how feedback has affected clump properties and whether the observed star-formation has been triggered.
   
    \par A broad classification of the effects of feedback on star-formation can lead to three different scenarios which we outline here:
    \begin {enumerate}
        \item Redistribution: The expansion of an HII region simply moves the star forming clumps to its edge. The molecular cloud is already seeded with dense clumps that would collapse to form stars. An apparent enhancement  
        of star formation observed on the periphery of the HII region is not necessarily caused by the HII region. 
        However, overall star formation has not been enhanced. We call this the ``redistribution'' scenario.

        \item Dispersion: In high mass star forming regions 
        feedback is very strong and can simply disperse most of the molecular material via various mechanisms \citep[and references therein]{krumholz2014}, thereby suppressing the ability of the GMC to form stars. We call this the ``dispersion'' scenario.

        \item Triggering: In this case, the feedback from the stars actually enhances or induces star-formation in the natal cloud. Many mechanisms are known or hypothesized to be responsible for triggering, e.g. supernovae \citep[and references therein]{degeus1992,nagakura2009}, cloud-cloud collisions \citep[for a comprehensive review, see][]{fukui2021}, collect and collapse \citep{elmegreen1977, dale2007}, radiation driven implosion \citep{bertoldi1989, bertoldi1990, kessel2003, lee2007}. In G305, there has been an indication of the ``collect and collapse'' (C\&C) mechanism playing a role in triggering star formation \citep{hindson10}. The C\&C model was first proposed by \citet{elmegreen1977}. In this scenario, the expanding warm ionized gas sweeps up a shell of shocked cool neutral gas. This compressed and shocked layer may become gravitationally unstable along its surface on a ``long'' timescale. This process allows massive dense fragments to form, which quickly fragment in turn leading to the formation of a cluster of stars of roughly the same age \citep{whitworth1994, deharveng2005}. We will therefore concentrate our efforts on testing this specific triggering mechanism in G305.
    \end{enumerate}

    \par In this paper we aim to find out which of the three scenarios mentioned above is dominant. Based on observations of NH$_3$ inversion lines, \citet{hindson10} found the concentration of star formation tracers in G305 to be enhanced inside a clump that faces the ionizing sources. The analysis presented in \citep{hindson10} is consistent with a triggering scenario but, the evidence is circumstantial. In order to robustly determine which of the three scenarios presented above dominates, we need to study the statistical properties of clumps, especially those related to star forming properties of the clumps, e.g. their masses, luminosity to mass ratios, surface mass densities. If the feedback is simply moving the clumps around, there should not be any significant difference in the mass distribution function of the clumps in G305 when compared to the Galactic average. For the ``dispersion'' case we would expect to see fewer clumps and fewer massive clumps. And finally, if triggering is dominant, we expect to see more massive clumps when compared to the Galactic average and, in case of an increased star forming efficiency, a higher luminosity to mass ratio.
    
    We discuss the observations and methods of data reduction in Sec. \ref{sec:data}. In Sec. \ref{sec:ancillary}, we introduce the ancillary survey data used to aid our analysis. Sec. \ref{sec:dendro} and \ref{sec:clump_props} explains the details of the clump extraction method and how their properties were estimated. Sec. \ref{sec:feedback} explores the effect feedback has on certain clump properties and how this may support or reject different scenarios presented in this section. In Sec. \ref{sec:galaxy_compare} we compare the properties of G305 clumps with the average Galactic sample to provide support for one of the three scenarios. Finally, Sec. \ref{sec:conc} presents the summary of our findings.

\section{Observations and Data Reduction \label{sec:data}}

    The Atacama Pathfinder Experiment (APEX) \citep{APEX} telescope\footnote{\tiny This publication is based on data acquired with the Atacama Pathfinder EXperiment (APEX). APEX is a collaboration between the Max-Planck-Institut f\"{u}r Radioastronomie, the European Southern Observatory and the Onsala Space Observatory.} was used to observe the 1 squared degree region spanning a longitude range $305\degree<l<306\degree$ and latitude $-0.5\degree<b<0.5\degree$. The Large APEX sub-Millimeter Array (LAsMA) receiver \citep{APEX-arrays} was used to observe the $J=3-2$ transitions of \coa ($\nu_{\text{rest}}\sim345.796\,\si{\GHz}$) and \cob ($\nu_{\text{rest}}\sim 330.588\,\si{\GHz}$) simultaneously. 
    
    \par
\begin{figure}
        \centering
        \includegraphics[trim=0 0.75cm 0 0, clip, width=0.5\textwidth]{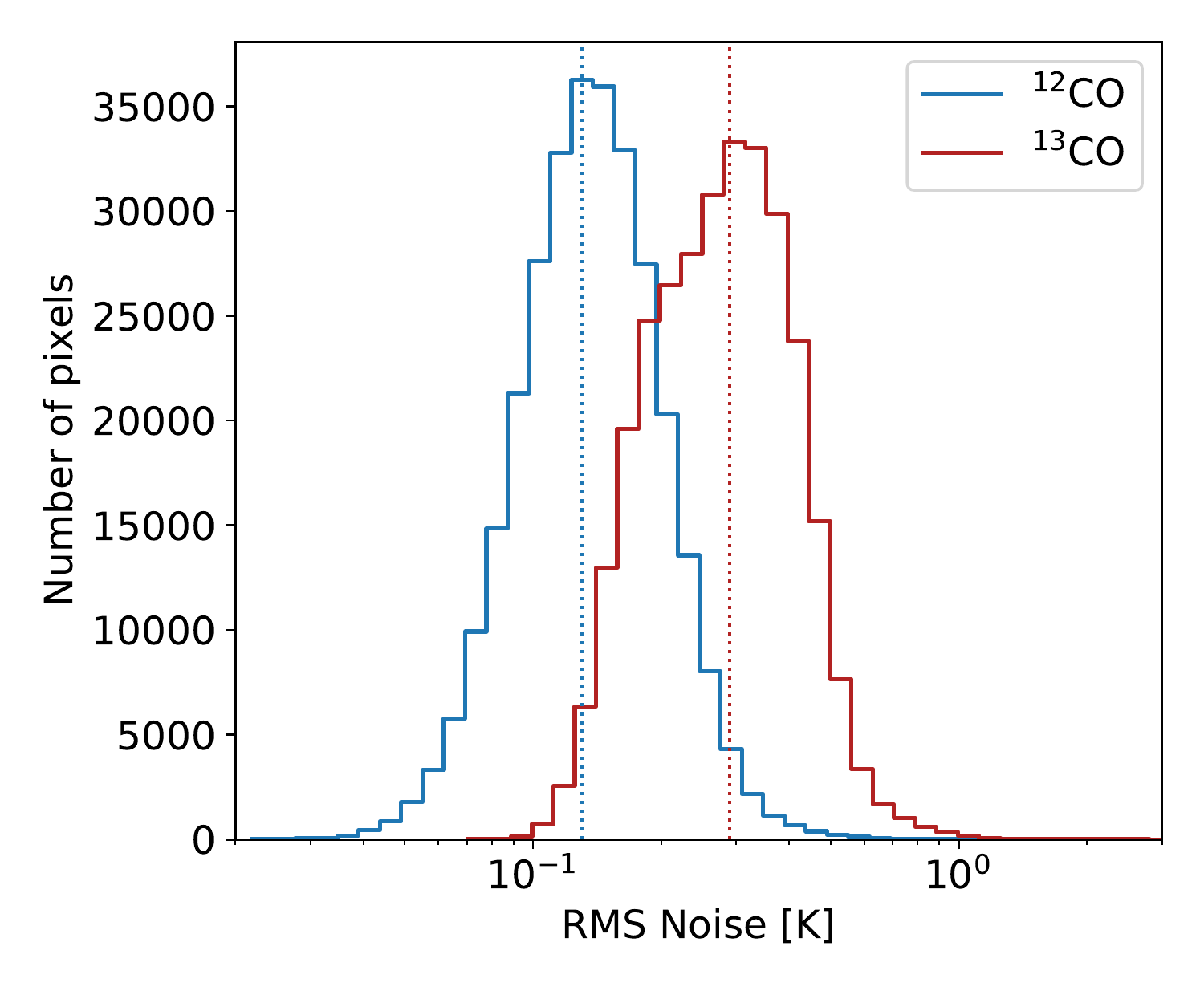}
        \caption{Histogram distribution of rms noise values for the LAsMA map of the G305 region. The \coa map's distribution peaks at 0.13\,K and that of \cob peaks at 0.29\,K as indicated by the blue and red dotted lines respectively.}
        \label{fig:noise_dist}
    \end{figure}

 The details of the observations, calibrations as well as data reduction has been described in \citetalias{mazumdar2021}. Here we briefly summarize the whole process. The local oscillator frequency was set at $\SI{338.190}{\GHz}$ in order to avoid contamination of the $\ce{^{13}CO}\,(3\text{-}2)$ lines due to bright $\ce{^{12}CO}\,(3\text{-}2)$ emission from the image band. The whole region was observed using a position switching on the fly (OTF) technique in perpendicular directions in order to avoid systematic scanning effects. The data was calibrated using a three load chopper wheel method, which is an extension of the ``standard'' method used for millimeter observations \citep{ulich} to calibrate the antenna temperature $T^*_A$ scale. The data was reduced using the GILDAS package\footnote{\url{http://www.iram.fr/IRAMFR/GILDAS}}. Baseline subtraction was done on each spectrum after masking the velocity range $\SI{-150}{\km\per\s}$ to $\SI{50}{\km\per\s}$ in order to avoid fitting emission features. The average of all the spectra was then used to mask the channels containing line emissions and baseline fitting was repeated to obtain a cleaner and flatter baseline. The resulting reduced spectra were then gridded with a $6\si{\arcsecond}$ cell (referred to as a pixel from now onward) and convolved with a Gaussian kernel with a full width at half maximum (FWHM) size of one-third the telescope FWHM beam width. The final datacube has a spatial resolution of $\sim \SI{20}{\arcsecond}$, and a velocity resolution of $\SI{0.5}{\km\per\second}$. In order to obtain the rms noise, the standard deviation was calculated over a range of 100 emission-free spectral channels for each pixel. Fig. \ref{fig:noise_dist} shows the distribution of the noise in the G305 region for both \coa and \cob lines. The noise distribution peaks at 0.13\,K for \coa and 0.29\,K for \cob. These values were therefore adopted to be the rms noise (referred to from now as $\sigma_{noise}$) for each map.
    
\section{Ancillary Data \label{sec:ancillary}}
\begin{figure}
    \centering
    \includegraphics[trim=0 0.3cm 0 0, clip,width=0.5\textwidth]{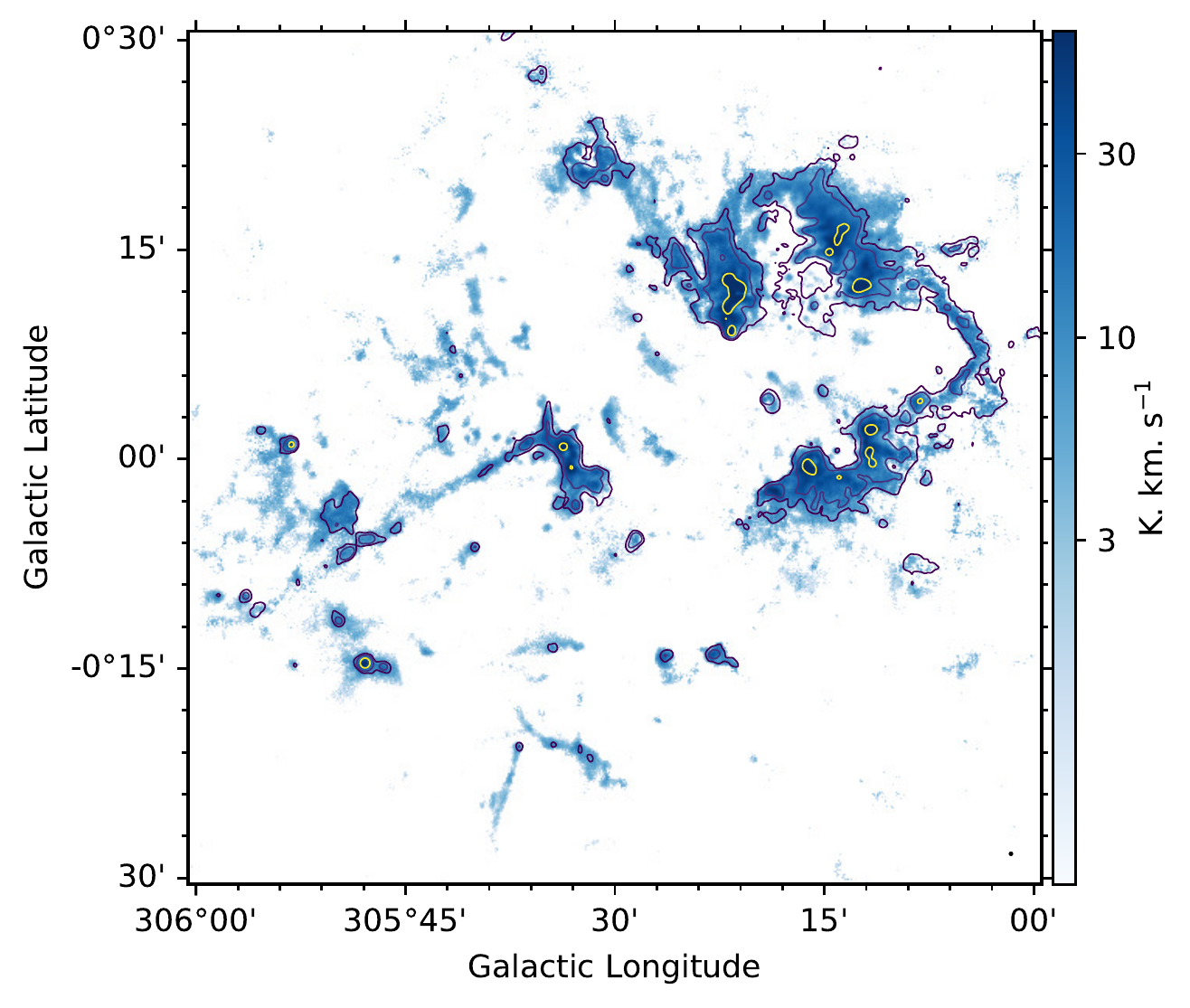}
    \caption{$^{13}$CO(3-2) moment-0 map of the G305 region. Overlaid on top are the contours of ATLASGAL corresponding to intensities 0.6, 1 and 3 Jy/beam.\label{fig:13CO_AGAL}}
    \end{figure}

    In order to compare and analyze G305 clump properties with the Galactic clumps, we have used the clump catalogs from the ATLASGAL \citep{agal_source, contreras2013, urquhart2014agal} and CHIMPS \citep{chimps_source} surveys. The Galactic clumps derived from the ATLASGAL survey \citep{agal_galactic_sample_2018} have the advantage of having a similar resolution as the LAsMA data (19\si{\arcsecond}). Additionally, there is a good agreement in the sensitivity between the ATLASGAL data and the \ce{^{13}CO} intensity map as can be seen in Fig. \ref{fig:13CO_AGAL}. The sensitivity to column density of both the data sets are also very similar with LAsMA tracing column densities down to $\rm{log}(N(H_2)[\rm{cm}^{-2}] \sim 21.46$ \citepalias[see][]{mazumdar2021} and ATLASGAL being sensitive to clumps with column densities down to $\rm{log}(N(H_2)[\rm{cm}^{-2}] \sim 21.5$ \citep{agal_source, agal_galactic_sample_2018}. On the other hand, CHIMPS survey uses the same spectral line ($^{13}$CO(3-2)) used in our survey, which makes comparison of properties more straightforward. But the clumps extraction from CHIMPS survey has been done on a smoothed dataset  of resolution $\sim \, 27\, \si{\arcsecond}$ \citep{rigby19}. Hence, in order to compare the properties of the G305 clumps with the CHIMPS Galactic clumps, we created an additional smoothed datacube and repeated all the analysis that follows in parallel. While creating this new dendrogram, a structure was now considered independent only if it were at least the size of the new beam. All comparison of properties with the CHIMPS clumps shown in this paper are based on the smoothed datacube.
    
    \par In order to compare the properties of G305 clumps with any Galactic sample, similar spatial scales are more important than similar angular resolution. Therefore, a distance limited sample ($3.5 < d < 4.5 \,\rm{kpc}$) of Galactic clumps has been used for comparison with clumps in G305. All references to the Galactic sample of clumps in this paper implicitly imply this distance limited sample. For the CHIMPS survey it is possible to choose a different range of distances to avoid smoothing our dataset in order to compare similar clump sizes. But that leads to fewer clumps in the CHIMPS sample (factor 1.5 fewer clumps). Therefore, the former method was adopted for comparison with the CHIMPS sample.
    
\section{Extracting clumps: Dendrogram \label{sec:dendro}}

        \begin{figure*}[t!]
        \centering
        \includegraphics[trim=0 0.7cm 0 0, clip,width=\textwidth]{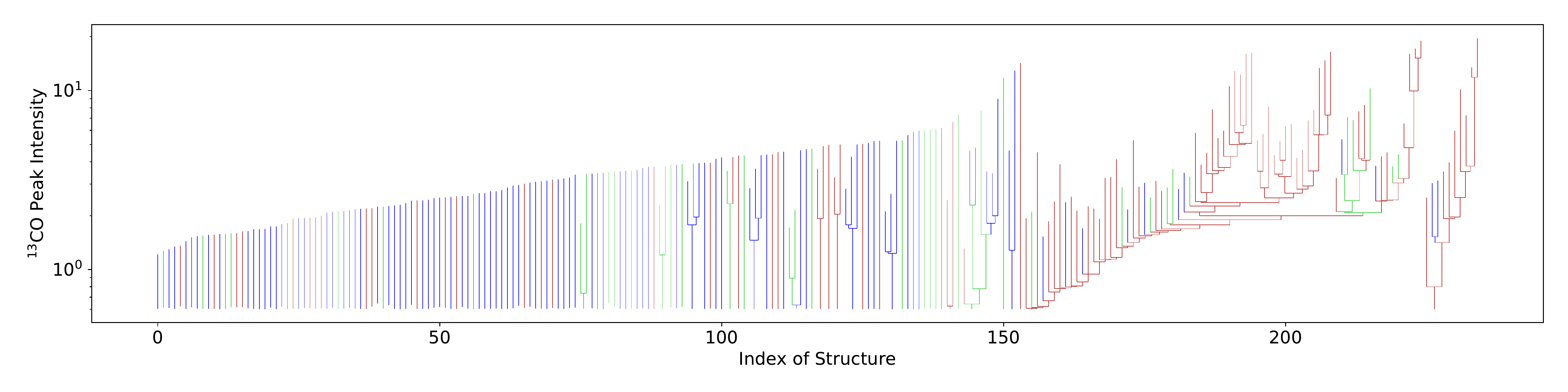}
        \caption{Dendrogram of G305 on \ce{^{13}CO} spectral cube. The structures that show greater than $\sim 67\%$ overlap with the 8\,\mum mask (regions with 8\,\mum flux greater than 100\,MJy/sr representing molecular gas influenced by feedback from the central star clusters in G305.) are colored red; those with partial overlap (between 10 - 67\%) are colored green; and the structures with less than 10\% overlap with the 8$\,\mu$m mask are colored blue.}   \label{fig:dendro13}
        \end{figure*}

        \par The clump catalog was created using the dendrogram analysis developed by \cite{dendrogram}. It decomposes the intensity data cube into a nested hierarchy of structures called leaves, branches and trunks. The leaves are the highest lying structures (with the brightest intensity) in this hierarchy and they generally represent the resolution element achievable by the survey designs. They consist of a collection of isocontours that contain only one local maximum within them. Branches are the structures that contain other leaves and branches inside them and trunks are the largest contiguous structures at the bottom of this hierarchy. Trunks by definition could also be leaves that are single isolated structures without any parent structure. Hence each structure identified by \texttt{astrodendro} as either a leaf or a branch depending on whether it has a lower hierarchy of structures under it.

        \par The dendrogram algorithm \texttt{astrodendro}\footnote{astrodendro is a Python package to compute dendrograms of Astronomical data (\url{http://www.dendrograms.org/}).} was run on the data cube in position-position-velocity (ppv) space. A lower limit ($min\_value$) of 5$\sigma_{noise}$ was set to avoid getting structures with peak intensity below the noise threshold. Additionally, a structure was considered independent only if its peak intensity differed by at least 5$\sigma_{noise}$ from the nearest local maximum ($min\_delta$). Each structure was required to have an area equal to area of the beam to be considered real. In the velocity space a minimum width of 4 channels ($=2\,\si{\km\per\s}$) was implemented to prune (get rid of unwanted structures due to noise in the data) the dendrogram. Figure \ref{fig:dendro13} shows the resulting dendrogram of the \cob data cube. The branching shows the fragmentation of a cloud into smaller and denser clumps.

\section{Catalog of clump properties \label{sec:clump_props}}

        \begin{figure*}[ht!]
            \centering
            \includegraphics[trim=0 0.7cm 0 0.3, clip, width=\textwidth]{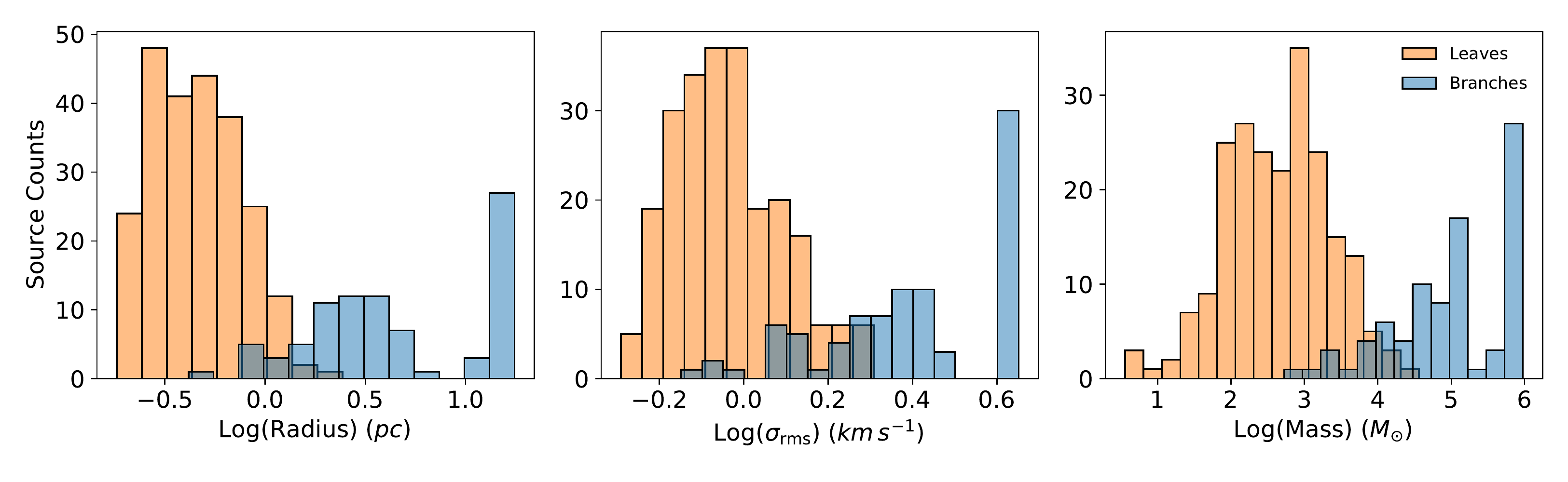}
            \caption{Comparison of radii (left), velocity dispersions (center), and masses (right) of leaves and branches in G305. \label{fig:structure_props}}
        \end{figure*}

        \begin{table*}[h!]
        \caption{Properties of the 20 randomly selected structures (leaves and branches) derived from the \cob dendrogram of G305. This table is meant to display a small subset of the whole catalog.\tablefootmark{a}}
        \label{tab:clump_props}
        \centering
        \begin{tabular}{ccccccccccc}
        \hline
        Catalog & Structure & x$_c$ & y$_c$ & Radius & v$_{\rm{rms}}$ & log(Lum) & log(Mass) & $\alpha_{\rm{vir}}$ & log($n(H_2)$) & log($\bar{F}_{8\,\mu m}$) \\
        Index & Type & & & & & & & & & \\
        & &($\mathrm{{}^{\circ}}$) & ($\mathrm{{}^{\circ}}$) & ($\mathrm{pc}$) & ($\mathrm{km\,s^{-1}}$) & ($\mathrm{K\,km\,s^{-1}\,pc^{2}}$) & ($\mathrm{M_{\odot}}$) &  & ($\mathrm{cm^{-3}}$) & ($\mathrm{MJy/sr}$) \\
        \hline \\
        18 & branch & 305.284 & 0.161 & 13.8 & 4.0 & 3.54 & 5.77 & 0.26 & 2.9 & 2.36 \\
        24 & branch & 305.804 & -0.235 & 2.5 & 1.9 & 2.18 & 4.37 & 0.25 & 3.76 & 1.74 \\
        28 & branch & 305.222 & 0.119 & 10.7 & 4.5 & 3.3 & 5.54 & 0.43 & 3.0 & 2.39 \\
        53 & branch & 305.284 & 0.161 & 13.8 & 4.0 & 3.54 & 5.77 & 0.26 & 2.91 & 2.36 \\
        60 & branch & 305.246 & -0.014 & 2.5 & 1.9 & 2.48 & 4.72 & 0.12 & 4.09 & 2.52 \\
        67 & leaf & 305.231 & -0.025 & 0.7 & 0.7 & 1.52 & 3.77 & 0.05 & 4.7 & 2.35 \\
        84 & branch & 305.865 & -0.045 & 4.2 & 1.4 & 2.43 & 4.54 & 0.17 & 3.21 & 2.02 \\
        86 & leaf & 305.612 & 0.143 & 0.8 & 1.8 & 0.99 & 3.69 & 0.35 & 4.58 & 1.8 \\
        116 & leaf & 305.175 & 0.254 & 0.5 & 0.9 & 0.01 & 1.66 & 5.67 & 3.21 & 2.15 \\
        119 & leaf & 305.223 & 0.324 & 0.6 & 1.0 & 0.54 & 2.68 & 0.82 & 3.95 & 2.06 \\
        123 & leaf & 305.928 & -0.1 & 0.3 & 1.2 & 0.05 & 2.02 & 2.35 & 4.33 & 1.7 \\
        129 & leaf & 305.208 & 0.141 & 0.6 & 0.7 & 0.72 & 2.88 & 0.31 & 4.09 & 2.1 \\
        137 & branch & 305.576 & -0.006 & 4.4 & 2.1 & 2.72 & 4.91 & 0.17 & 3.53 & 2.4 \\
        159 & branch & 305.361 & 0.2 & 1.0 & 2.3 & 1.99 & 4.23 & 0.21 & 4.81 & 2.99 \\
        187 & branch & 305.564 & 0.002 & 2.3 & 1.8 & 2.42 & 4.61 & 0.12 & 4.09 & 2.53 \\
        249 & leaf & 305.67 & 0.025 & 0.6 & 1.5 & 0.71 & 2.74 & 1.62 & 3.95 & 2.29 \\
        267 & leaf & 305.552 & -0.055 & 0.7 & 1.2 & 1.16 & 3.47 & 0.23 & 4.49 & 2.29 \\
        272 & leaf & 305.681 & 0.307 & 0.9 & 0.9 & 1.14 & 3.65 & 0.13 & 4.3 & 1.91 \\
        292 & leaf & 305.493 & 0.01 & 0.3 & 0.8 & -0.28 & 1.84 & 1.71 & 4.11 & 1.82 \\
        321 & leaf & 305.28 & -0.153 & 0.6 & 1.0 & 0.32 & 1.86 & 5.77 & 3.1 & 1.76 \\

        \hline
        \end{tabular}
        \tablefoottext{a}{The full table has been made available electronically.}
        \end{table*}

        Once the dendrogram was created, the \texttt{astrodendro} package was used to create a catalog of some basic properties of all the structures. These include their positions, mean and RMS velocities, total fluxes and the projections of their major and minor axes (called $\sigma_{\text{maj}}$ and $\sigma_{\text{min}}$) on the position-position plane \citep{rosolowsky2006}. The radius of the clumps is defined by $R_{\text{eq}} = \eta*\sqrt{\sigma_{\text{maj}}*\sigma_{\text{min}}}$, where $\eta$ is a factor that relates the radius of a spherical cloud to its one dimensional RMS size $\sqrt{\sigma_{\text{maj}}*\sigma_{\text{min}}}$. For consistency with \citet{solomon1987,rosolowsky2006,colombo2019,rigby19} we adopt the value $\eta=1.91$. The velocity dispersion ($\sigma_v$) is calculated as the intensity weighted second moment of the velocity axis. The flux of the region is the sum (zeroth moment) of all the emission in the region. The physical radius of the clumps and their luminosities were calculated using $R_{\text{pc}}=Rd$ and $L=Fd^2$ respectively, where $d=3.8\,\si{kpc}$ is the distance to G305 \citep{clark,borissova}.
        
        \par Masses of the clumps were calculated using the column density maps. Sect. 4 of \citetalias{mazumdar2021} describes in detail how the column density map was obtained for the region. We present a brief summary of the method here. Assuming that \ce{^{12}CO} is optically thick, the excitation temperature $T_{ex}$ of each pixel in the position-position-velocity space (referred to from now onward as voxel) was calculated under the assumption of local thermodynamic equilibrium (LTE). Then, the \ce{^{13}CO} optical depth was derived for each voxel from the excitation temperature obtained from $^{12}$CO and the $^{13}$CO intensity. Subsequently, the \ce{^{13}CO} column density for each voxel was calculated using the derived excitation temperature and optical depth. Since, the column density derived is only available for those pixels with a significant emission in \cob, all the properties calculated (and consequently all discussions about the clump properties) from here on are only applicable to the clumps extracted from the dendrogram of the \cob map of the region.
        As mentioned in Sec. \ref{sec:ancillary} this has the added advantage of having a similar sensitivity to ATLASGAL data as well as using the same transition as the CHIMPS survey.

        \par The column density over all the voxels corresponding to each clump in the dendrogram were added and the mass of the structure was then calculated using the equation:

        \begin{equation}\label{eq:1}
         M = \mu \, m_p \, R_{13}^{-1} \, \sum_{lmv}N_{13}(\rm{total})_{lbv} \, \Delta x \Delta y
        \end{equation}

        where $\mu$ is the mean mass per H$_2$ molecule, taken to be 2.72, accounting for a helium fraction of 0.25 (\texttt{cite Allen 1973}), $m_p$ is the mass of a proton, $R_{13}^{-1}$ is the abundance ratio of H$_2$ compared with $^{13}$CO, and $\Delta x$ and $\Delta y$ are the resolution element in parsec calculated using the distance of the source and the resolution element. The conversion $R_{13}^{-1}$ is calculated in two steps, and we adopt a ratio of $R_{12} /R_{13}$ that varies as a function of Galactocentric distance as prescribed by \citet{milam2005}, and a value of $R_{12} = 8.5 \times 10^{-5}$ \citep{frerking} is adopted for all sources. $R_{12}/R_{13}$ has a value of approximately 59.7 for G305's Galactocentric distance of 6.6$\,\si{kpc}$.
        
        \par The masses of the clumps and their radii were also used to calculate their density($n(H_2)$) and surface mass densities using:
         \begin{align} \label{eq:2}
            n(H_2) &= 15.1 \times M / (4/3 \, \pi R_{\rm{eq}}^3) \\
            \Sigma &= M/\pi \, R_{\rm{eq}}^2
         \end{align}
        Here $M$ is in $\si{M_{\odot}}$, $R_{\mathrm{eq}}$ is in $\si{pc}$, $n(H_2)$ is in $\si{\per \cubic \cm}$, and $\Sigma$ is in $\si{M_{\odot}\per\square {pc}}$. The factor of 15.1 is to convert the density to appropriate units.

        \par The dynamical state of the clumps, i.e. whether they are gravitationally stable, collapsing or dissipating is examined using the virial theorem. With only gravitational forces the virial theorem reads $2K + \Omega = 0$ where $K$ is the kinetic energy and $\Omega$ is the gravitational energy. The virial parameter is defined as the ratio of the virial mass of a spherically symmetric cloud to its total mass as given below:
        
        \begin{equation}\label{eq:4}
            \alpha_{\rm{vir}} = \frac{3 \sigma_{\rm{v}}^2 R_{\rm{eq}}}{G M}.
        \end{equation}
        Here, $\sigma_{\rm{v}}$ is the velocity dispersion of the clump and $G$ is the gravitational constant. This is the definition used by \citet{maclaren1988}. We also assume a spherical density distribution $\rho (r) = 1/r^2$. In the absence of pressure supporting the cloud, $\alpha_{\rm{vir}} < 1$ means that the cloud is gravitationally unstable and collapsing whereas for $\alpha_{\rm{vir}} > 2$ means that the kinetic energy is higher than the gravitational energy and hence the cloud is dissipating. A value of $\alpha_{\rm{vir}}$ between 1 and 2 is interpreted to be an approximate equilibrium between the gravitational and kinetic energies. A cloud undergoing gravitational collapse can also show $\alpha_{\rm{vir}} \sim 2$ as the rapid infall can manifest as a large velocity dispersion \citep{kauffmann2013}.

        The dendrogram resulted in a total of 337 structures of which 235 (69.73\%) are leaves, 102 (30.27\%) are branches. Fig. \ref{fig:structure_props} shows properties of leaves and branches to demonstrate their differences. Table \ref{tab:clump_props} shows 20 randomly selected clumps in the G305 giant molecular cloud.

\section{Effect of feedback on G305 clumps\label{sec:feedback}}

    The radiation and ionization front from the OB stars in the central cavity can effect the masses as well as the velocity dispersion of the clumps in G305. It can inject turbulence which should manifest as broad linewidths of the clumps. It can also compress and destabilize the clumps resulting in an increase in their surface mass densities.
    
    \subsection{Insufficient velocity resolution}
      Before we investigate the effects of feedback on the linewidths of clumps we first calculate whether we posses enough resolution in velocity to discern any differences that might occur in their linewidths due to feedback. Assuming that the observed width of a clump is solely attributed to its turbulence, we can write the turbulent pressure in a clump as,
        \begin{equation} \label{eq:Pnt}
         P_{\rm{NT}} = n(\ce{H2})m_{\ce{H2}}\sigma^2 / k \, ,
        \end{equation}
        where $k$ is the Boltzmann's constant and $\sigma$ is its velocity dispersion in one-dimension. If all the turbulent pressure is being delivered to the clumps by the stars in the center of G305 then $P_{\rm{star}} = P_{\rm{NT}}$, where $P_{\rm{star}}$ is given by,
        \begin{equation} \label{eq:Pstar}
         P_{\rm{star}} = Q(H^0) \langle h\nu \rangle / 4 \pi d^2 c k \, ,
        \end{equation}
      where $Q(H^0)$ is the number of H ionizing photons per second emitted by the stars, $\langle h\nu \rangle$ is the mean photon energy of an O-star ($\sim 15\si{\eV}$ \cite{Pellegrini07}), $d$ is the distance of the star (cluster) from the cloud surface, and $c$ is the speed of light. It is tricky to calculate $P_{\rm{star}}$ given that the distance to each cloud is different and the stars are also not localized. But in order to obtain an upper limit on the value of $P_{\rm{star}}$ we assume the total ionizing radiation of Danks 1 and 2 combined. In Danks 1 we have 2 O8-B3 and 1 O8-B3I stars for which we adopt the value corresponding to 3 O8V stars from Table 2 of \citet{panagia1973}. Similarly, we use O4V values for the 3 O4-6 stars, O6V values for the O6-8 and 2 O6-8If stars. For the 3 WNLh stars we adopt the value corresponding to WN-3w from table 2 of \citet{crowther2007}. Therefore for Danks 1 we have a maximum ionizing flux equal to $ 3.67 \times 10^{50} \,\si{\per \s}.$ For Danks 2, there are 3 O6 (O6-8), 4 O8 (O8-9 / O8-B3I), and 1 WC7 (WC7-8) stars (we have ignored the F8-G1 star). Assuming all these belong to class V, we get $8.03 \times 10^{49} \,\si{\per \s}$. Hence, Danks 1 and 2 have a combined maximum Lyc photon output of $4.473 \times 10^{50} \, \si{\per \s}$. The value of $d$ is assumed to be $4\,\si{pc}$ which is approximately the closest distance of clumps in the northern and southern complex to the Danks cluster. Plugging these into Eq. \ref{eq:Pstar}, we obtain a value of $P_{\rm{star}} \sim \num{1.36e6} \, \si{\kelvin\per\cubic\cm}$. This corresponds to a velocity dispersion of $\sigma = 1.06\,\si{\km\per\s}$ given an \ce{H2} density of $10^4\,\si{\per\cubic\cm}$. Since our dataset has a velocity resolution of 0.5$\,\si{\km\per\s}$, it is difficult to discern the effects of feedback on the velocity dispersion of the clumps, which will realistically be much smaller than $\sim 1\,\si{\km\per\s}$. The linewidths of the clumps were plotted as a function of the 8\,\mum flux and no dependence was observed between the linewidths of the clumps at the highest hierarchical level (leaves) and the strength of the feedback on the clumps supporting our claim. These results have been presented as an appendix (Fig. \ref{fig:vrm_vs_8um}) as they do not provide any new information to the question we are trying to answer in this paper.

    \subsection{Surface mass densities of clumps vs 8\,\mum \label{sec:clumpv8um}}

        \begin{figure}[h!]
            \centering
            \includegraphics[width=0.5\textwidth]{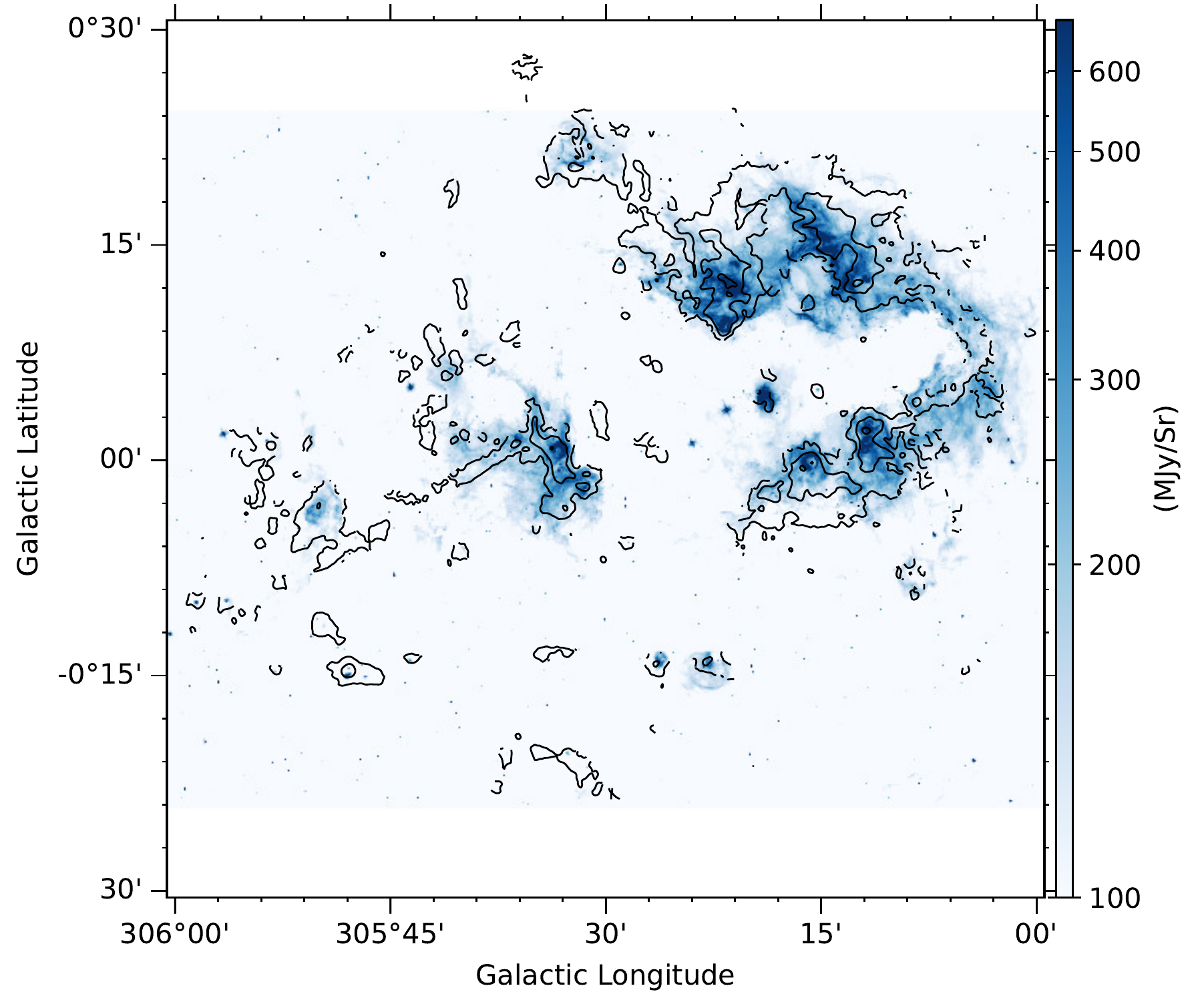}
            \caption{GLIMPSE $8\,\si{\micro\metre}$ map of the G305 regions. The black contours correspond to the $^{13}$CO J=3--2 integrated intensities of 5(5$\sigma$), 20(20$\sigma$) and 80 (80$\sigma$)$\,\si{\kelvin\km\per\second}$.}
            \label{fig:8um_vs_13CO_map}
        \end{figure}

        \par In \citetalias{mazumdar2021} we discussed the role of the Galactic Legacy Infrared Mid-Plane Survey Extraordinaire \citep[GLIMPSE:][]{Benjamin_glimpse,Churchwell_glimpse} 8$\,\mu$m continuum map as an indicator for feedback strength. As a proxy to the strength of feedback on each clump, we calculated the average 8$\,\mu$m flux over each clump. For this, a mask was extracted for each structure by projecting it on to the position-position plane and the 8$\,\mu$m flux was averaged over the mask. Table \ref{tab:clump_props} also shows the mean 8\,\mum flux associated with each structure.
        
        \par A significant issue with this analysis is that the 8\,\mum map is a 2 dimensional projection on the plane of G305 and hence it is possible for clumps to appear to be effected by feedback just because of their projection. In order to test whether the associated 8\,\mum fluxes actually correspond to the clumps we first compared the morphology of the 8\,\mum emission with the \cob integrated intensity map. Fig. \ref{fig:8um_vs_13CO_map} shows the GLIMPSE map with the \cob integrated intensity contours overlaid on top. There exists very good correlation between the two. So, we safely assume here that all the 8\,\mum emission is originating in G305. We then investigated whether any leaves in the dendrogram had the same position but different velocities along the line of sight. We found only about 1\% of the leaves ($\sim 7\%$ if branches are included) had positions within one beam size distance of another leaf (or branch).  Additionally, \citet{hindson13} looked at the morphology of the molecular gas emission and argued that the G305 complex has a flattened geometry instead of a spherical structure. Taking these factors into consideration, one can safely ignore the effects of projection on the associated 8\,\mum fluxes for the clumps.

        \begin{figure}[h!]
           \centering
           \includegraphics[width=0.5 \textwidth]{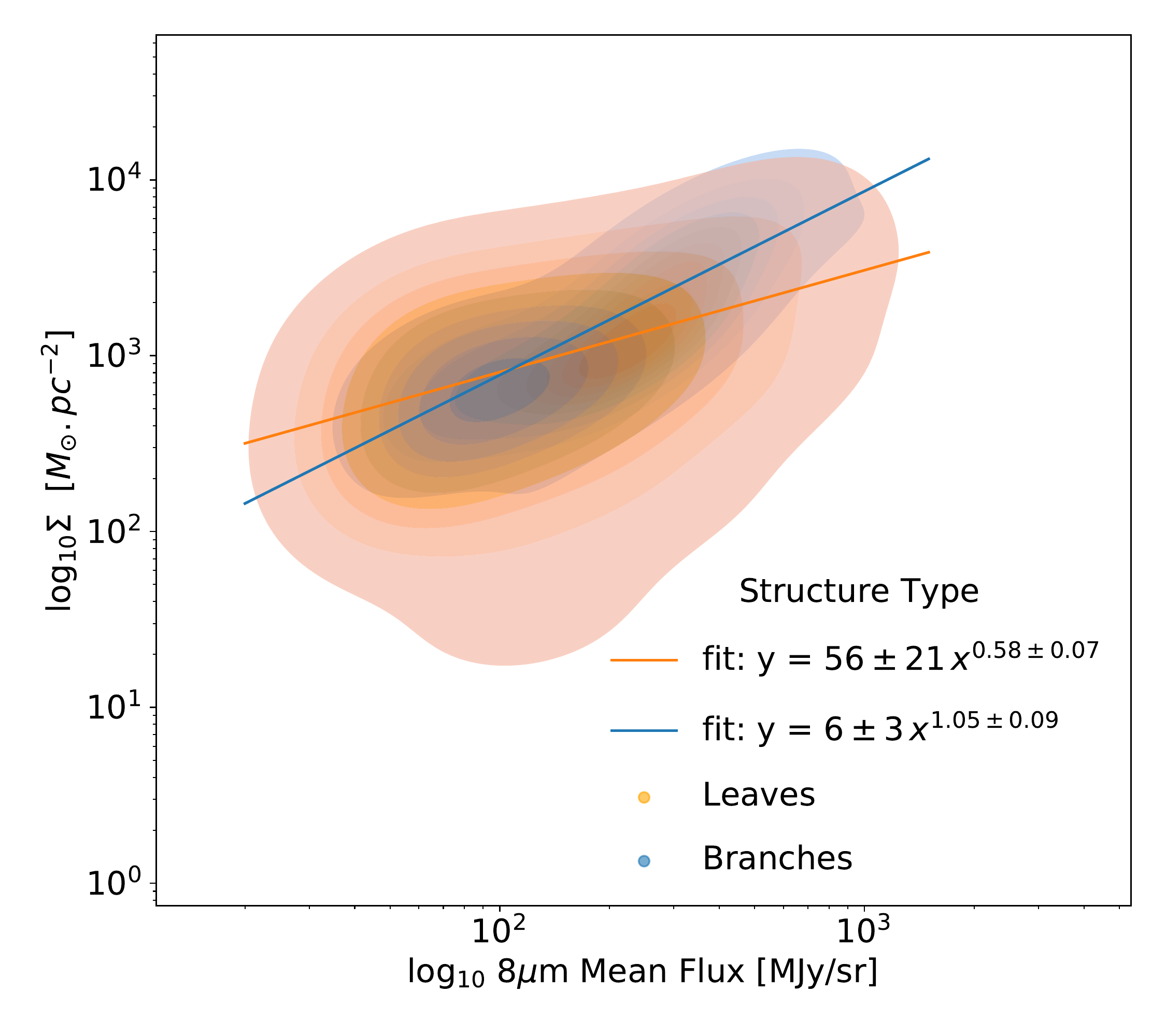}
           \caption{Surface mass density of clumps in G305 as a function of 8$\mu$m flux. The orange and blue lines show the results of a power law fit to these quantities for leaves and branches respectively. \label{fig:sigma_vs_8um}}
        \end{figure}

        \par Fig. \ref{fig:sigma_vs_8um} shows the two-dimensional kernel density estimate (KDE) distribution of the clump surface mass density as a function of 8$\,\mu$m flux. The leaves and branches of the dendrogram have been plotted separately. We observe that the surface mass density of the clumps increase with increasing 8$\,\mu$m flux indicating that in regions of stronger feedback the clumps are more massive. A power law fit between the two quantities resulted in the following relation between them for leaves and branches.
        \begin{align}
         \Sigma_l & = 56\pm21 \, {\rm{F}}_{8\mu m}^{\,0.58\pm0.07} \\
         \Sigma_b & = 6\pm3 \, {\rm{F}}_{8\mu m}^{\,1.05\pm0.09} \, ,
        \end{align}
        where $\Sigma_l$ and $\Sigma_b$ are the surface mass densities of the leaves and branches respectively, and $\rm{F}_{8\mu m}$ is the average 8\mum flux incident on the structure. A positive correlation between the surface mass density of the clumps and the incident 8\,\mum flux is indicative of triggering in G305. In the ``redistribution'' scenario we would have expected no correlation between the two quantities and in case of the ``disruptive'' scenario one should observe a negative correlation.

  \subsection{Inside and outside the feedback zone}

        \par The analysis described in the previous subsection motivated us to follow an approach similar to that presented in \citetalias{mazumdar2021}. Based on an 8\,\mum threshold a ``feedback zone'' was defined as the pixels with 8$\,\mu$m flux larger than the threshold. The value of this threshold was chosen to be $100\,\si{MJy \per sr}$ corresponding to the first inflection point in Fig. 15 of \citetalias{mazumdar2021}. We believe this contour corresponds to the isosurface where the stellar feedback gets deposited first. Then each clump was assigned a tag depending on the extent of its overlap with the ``feedback zone''. Clumps with more than 67\% overlap with the ``feedback zone''  were tagged ``Mostly Inside''; those with an overlap between 10 and 67\% were tagged ``Partly Inside'', and those with less than 10\% overlap with the ``feedback zone'' were tagged ``Outside''. This classification is not based on any physics but in order to aid the explanation that is to follow. Using this classification results in 95 leaves (40.4\%) that were ``mostly inside'', 40 leaves (17.0\%) were ``partly inside'', and 100 leaves (42.6\%) were ``outside'' the feedback zone. The same was done with branches which led to 75 branches (73.6\%) ``mostly inside'', 14 (13.7\%) ``partly inside'', and 13 (12.7\%) ``outside'' the feedback zone.
       
       \begin{figure}[h!]
         \centering
         \includegraphics[width=0.5 \textwidth]{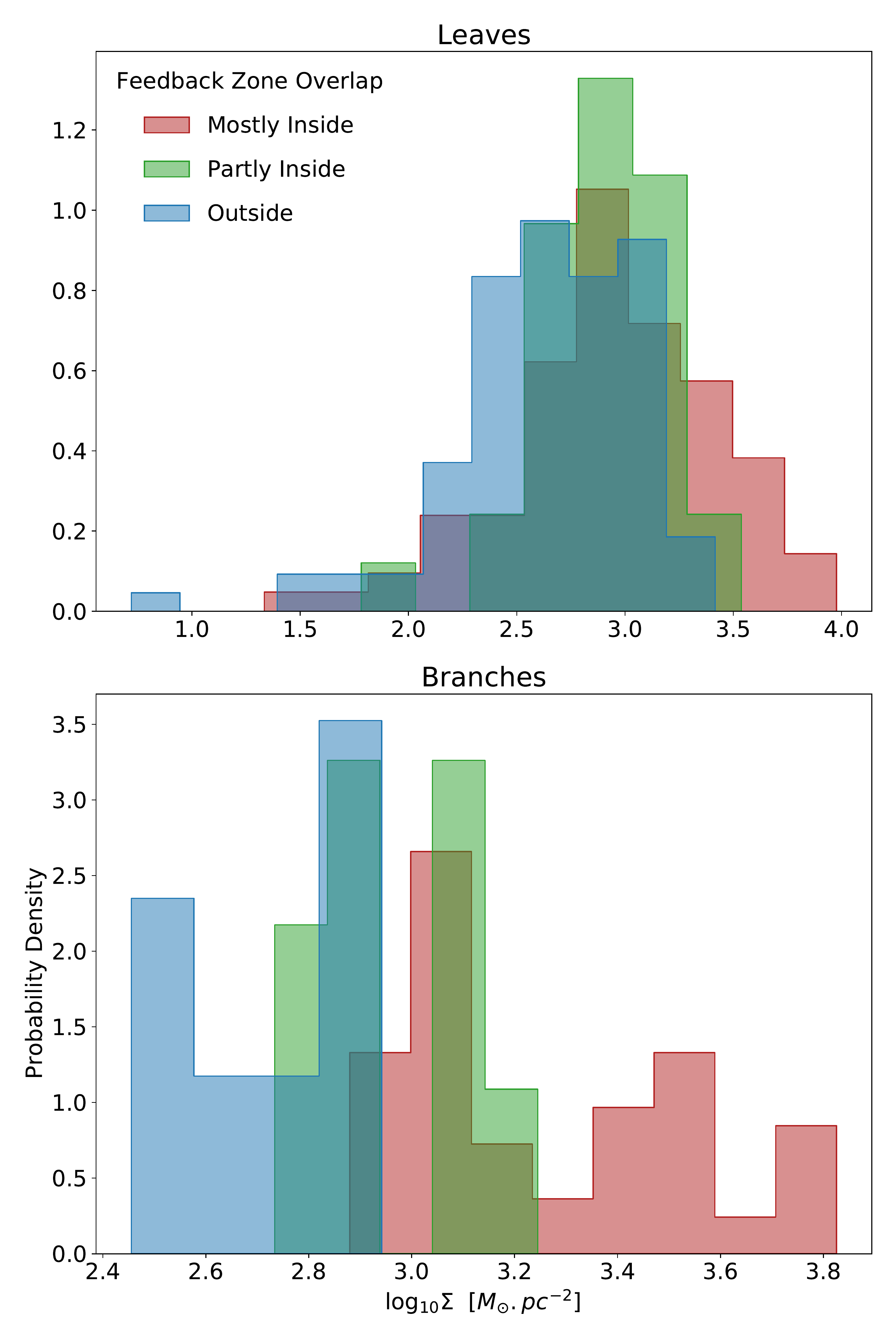}
         \caption{Histogram of the probability density of surface mass densities of leaves (top) and branches (bottom) based on their overlap with the ``feedback zone'' defined using an 8\,\mum threshold of $100\,\si{MJy \per sr}$.}
         \label{fig:sigma_of_tags}
        \end{figure}

        \par Fig. \ref{fig:sigma_of_tags} shows the probability density distribution of the surface mass densities of the structures based on their tags. There is a clear trend of increasing surface mass density in association with the overlap with the feedback zone. A Kolmogorov-Smirnov (KS) test \citep{ks-test-1,ks-test-2} was performed on the pairs of these distributions to test the hypothesis that these samples of surface mass densities have been taken from the same population. In the case of leaves only the ``mostly inside'' and the ``outside'' samples rejected the null hypothesis (p-value = 0.0011\%). The other pair of samples did not reject the null hypothesis (p-value = 0.12 for ``mostly inside'' vs ``partly inside'' and 0.18 for ``partly inside'' vs ``outside''). But the branches in the ``mostly inside'' category were significantly different from ``mostly outside" one (p-value$<<0.0013$) but not from the ``partly inside" ones (p-value=0.002). These findings supports the collect and collapse model that predicts the clumps to be more massive under feedback.

        \par Another prediction of the C\&C model is that there will be a higher level of fragmentation in the cloud being affected by feedback. To test this, we color coded the structures of the dendrogram in Fig. \ref{fig:dendro13} based on their tags. We observe the largest hierarchical structure almost completely lies ``mostly inside'' the feedback zone with some structures being ``partly inside''. The structures outside the feedback zone show very little fragmentation. This is further support for the C\&C model in the case of G305.

\section{G305 vs Galactic clumps \label{sec:galaxy_compare}}
        \par We have seen so far that feedback from the central cluster of stars has a significant impact on the clumps in G305. This suggests that the properties of the clumps in G305 should be significantly different from the average sample of Galactic clumps. In order to test this we compared the properties of G305 clumps with those of the Galactic sample of clumps derived from ATLASGAL and CHIMPS.
        \begin{figure*}[h!]
         \centering
         \includegraphics[width=\textwidth]{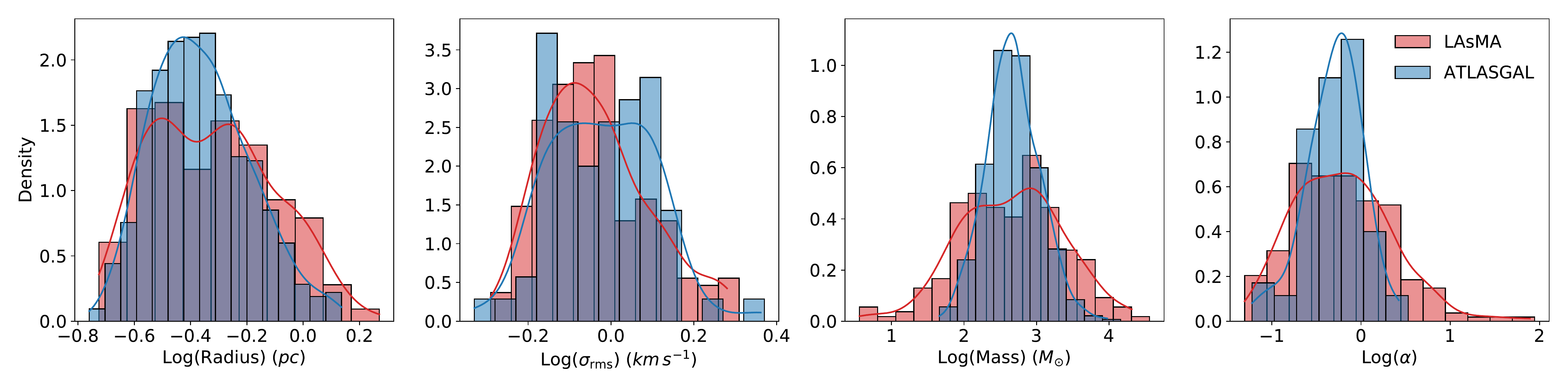} \\
         \includegraphics[width=\textwidth]{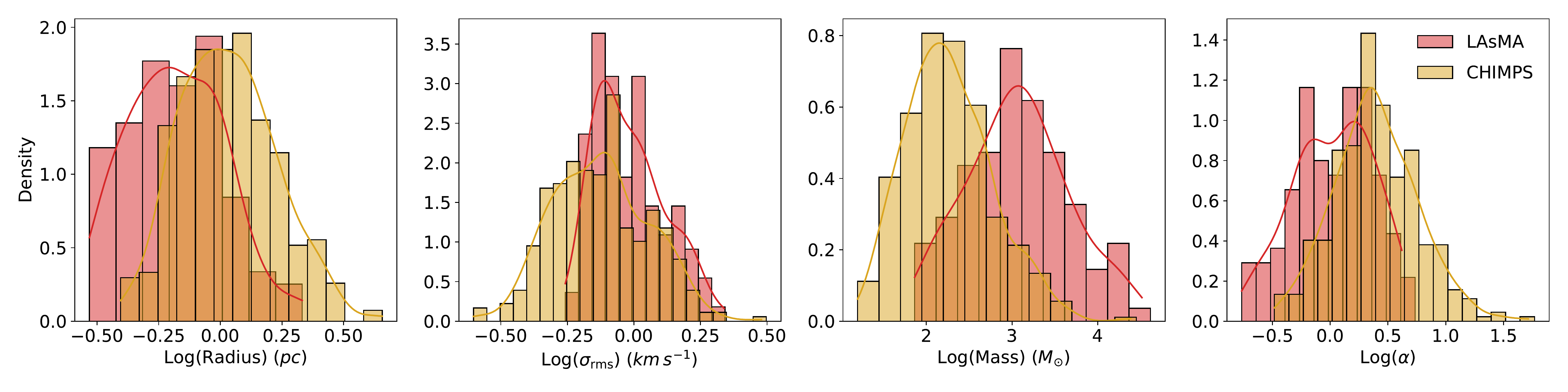}
         \caption{Comparison of properties of LAsMA clumps in G305 with a Galactic sample derived from ATLASGAL (top panel) and CHIMPS (bottom panel) surveys. The solid lines overlaid on top in matching colors show the KDE of the distribution of the corresponding properties.}
         \label{fig:Clump_Compare}
        \end{figure*}

        Fig. \ref{fig:Clump_Compare} shows the distribution of different properties of clumps in G305 along with the Galactic sample derived for both ATLASGAL clumps \citep{agal_galactic_sample_2018,contreras2013} and CHIMPS clumps \citep{rigby19}. In order to compare the clumps properties only the CHIMPS clumps with the highest reliability flag (= 3) were considered \citep{rigby19}.
        \par The radius of the clumps in G305 are smaller than the CHIMPS clumps. This could be because of the differences in the clump extraction method used by \citep{rigby19}. Compared to the dendrogram clumps, the \texttt{FELL-WALKER} method \citep{fellwalker} would have slightly larger areas. In comparison to ATLASGAL, the G305 clumps appear to have a different distribution. But this difference is not significant as confirmed by performing a KS-test on the radius of the two populations (p-value=0.0022).
        \par The linewidths of the G305 clumps appear to have a different distribution from the ATLASGAL and CHIMPS samples. A KS-test performed on the linewidths of G305 clumps with the two Galactic samples resulted in a p-value$<<0.0013$ for both the populations confirming this. The ATLASGAL survey used the NH$_3$ spectral line to estimate linewidths whereas the CHIMPS survey used the same line used in our observations and hence this comparison is more straightforward. We observe the linewidths to be higher in case of G305 clumps when compared to the CHIMPS population of Galactic clumps.
        
        \par The G305 clumps show a much broader distribution of masses compared to ATLASGAL. Higher sensitivity of our observations would explain why we see more clumps of smaller masses as compared to ATLASGAL as these would not be detected in the ancillary survey. In comparison to the CHIMPS population of clumps the masses of G305 clumps are much higher. G305 also has a higher fraction of heavier clumps compared to the ATLASGAL sample. This is indicative of some form of triggering. A KS test on the masses of the two populations resulted in a p-value << 0.0013 for masses of G305 clumps as compared to the Galactic sample (for both ATLASGAL and CHIMPS populations) indicating that the G305 clumps are significantly different from the Galactic average.
        
        \par The last panel in Fig. \ref{fig:Clump_Compare} shows the distribution of virial parameters of the G305 clumps. We see a broader distribution when compared to ATLASGAL clumps. A large proportion of clumps in G305 appear to have $\alpha_{\rm{vir}} << 2$. According to \citet{kauffmann2013} this is indicative of high-mass star formation in G305. We also see a much higher fraction of clumps in G305 with $\alpha_{\rm{vir}} >> 2$ which may be indicative of turbulence disrupting and dissipating clumps. Interestingly, the smoothed dendrogram does not have clumps with very high $\alpha_{\rm{vir}}$ values. This could be attributed to the bias towards larger structures in the smoothed dataset leading to more clumps with heavier masses. A KS-test on the virial parameters of the G305 and Galactic clumps indicated these properties being drawn from significantly different populations (p-value$<<0.0013$).

    \subsection*{Triggering in G305 \label{sec:trig}}

        Proving that triggering is responsible for enhanced star formation is a difficult task. Often it is done by looking at the morphological signposts like star-formation at the rim of an expanding shells of HII regions \citep{deharveng2005, zavagno2005, urquhart2007, thompson2012, kendrew2012, palmeirim2017}. In G305, \citet {hindson13} looked at star-formation in different phases and compared their distribution in the complex to suggest that the most likely case of triggering is the collect and collapse process. However, they did not find any conclusive evidence that the star-formation happening in G305 has been triggered. They calculated the time needed for the gas to fragment in to stars and suggested that while in some cases the age of the triggering stars was old enough to be responsible for the stars in their vicinity to be triggered. In many other cases, the stars responsible for triggering may not be old enough to drive the formation of the HII and UC HII regions. But the fragmentation time-scale estimates were based on assumptions about the initial ambient atomic density and the Lyman continuum photon rate both of which are difficult to measure, making their suggestions inconclusive. So far in this paper, we have explored whether the feedback from the stars is indeed triggering star formation in G305 by looking at their impact on the clumps which are the reservoirs of the new generation of stars.
        \begin{figure}[h!]
            \centering
            \includegraphics[width=0.4 \textwidth]{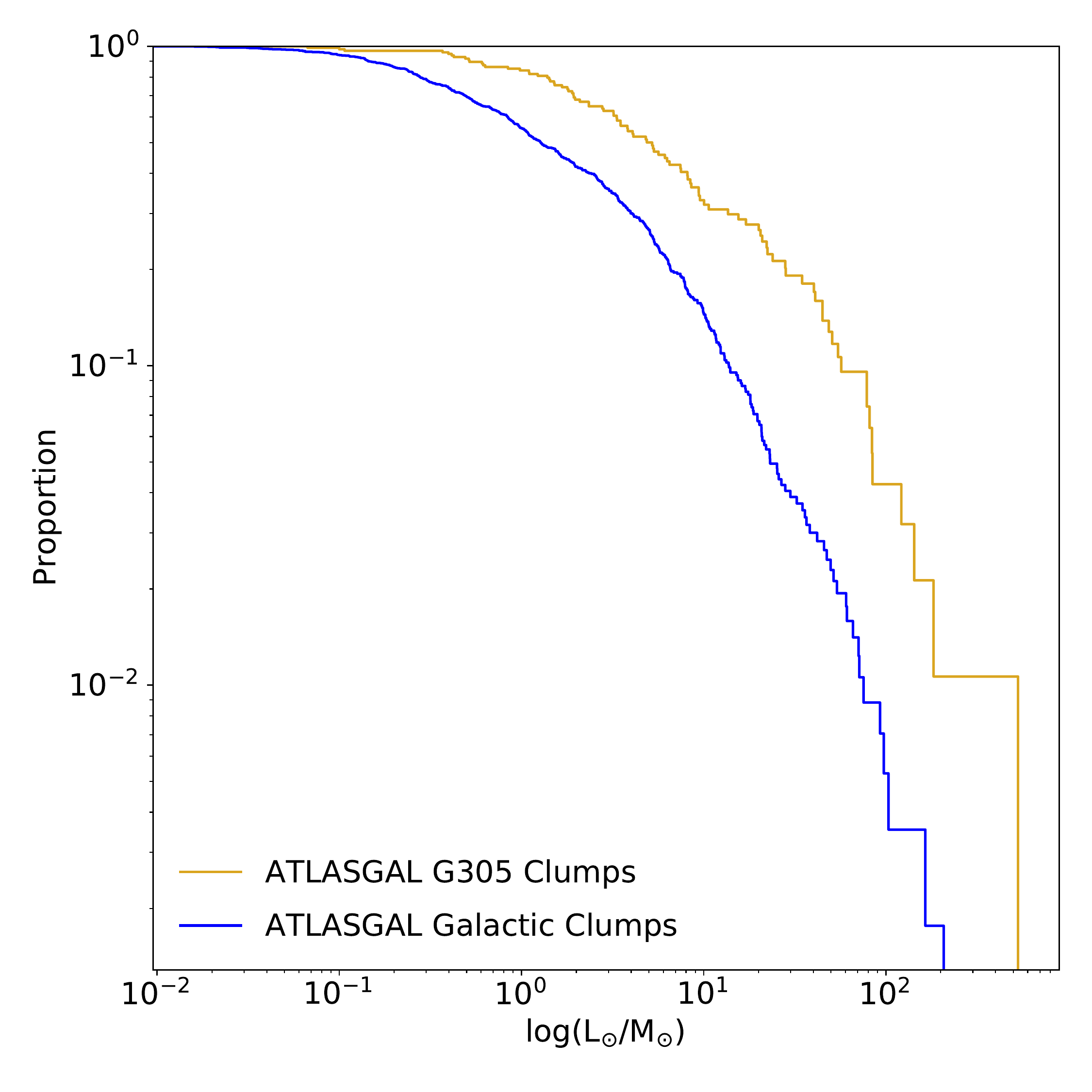}
            \caption{\texttt{[Sec.4]:} CDF of $L_{clumps}/M_{clumps}$ in G305 compared to that of the ATLASGAL Galactic clumps. \label{fig:L/M}}
        \end{figure}

        The luminosity to mass ratio ($L/M$) of clumps is a good indicator of the evolutionary stage in star formation \citep[see][]{molinari2008,agal_galactic_sample_2018}. A higher star forming efficiency would mean that more clumps are in more evolved stages of forming stars leading to a flatter cumulative distribution function of their $L/M$. In G305, if the clumps are simply being moved around without any effect on their star forming efficiency, then the $L/M$ distribution of the clumps in G305 should not vary significantly from the Galactic average. Fig. \ref{fig:L/M} shows the cumulative distribution function (CDF) of the $L/M$ for the ATLASGAL clumps in G305 compared to that of the the overall ATLASGAL Galactic sample. All the $L/M$ values were taken from ATLASGAL as the calculation of the luminosity is quite different between our dataset and ATLASGAL. Consequently, comparing the $L/M$ of clumps from our data with that of the ATLASGAL clumps would not have been straightforward. The clumps in G305 show significantly higher $L/M$ than the Galactic average as is shown by the CDF. A KS test on the two clump samples returned a p-value$<<0.0013$. Therefore, the null hypothesis is quite overwhelmingly rejected indicating that the clumps in G305 show higher star forming efficiency compared to the Galactic average. This rules out the ``redistribution'' and ``disruption'' scenarios. These clumps may not have a one to one correspondence with the \cob clumps derived in the previous sections, but that does not negate the validity of the finding here.

        \begin{figure}[h]
            \centering
            \includegraphics[width=0.4\textwidth]{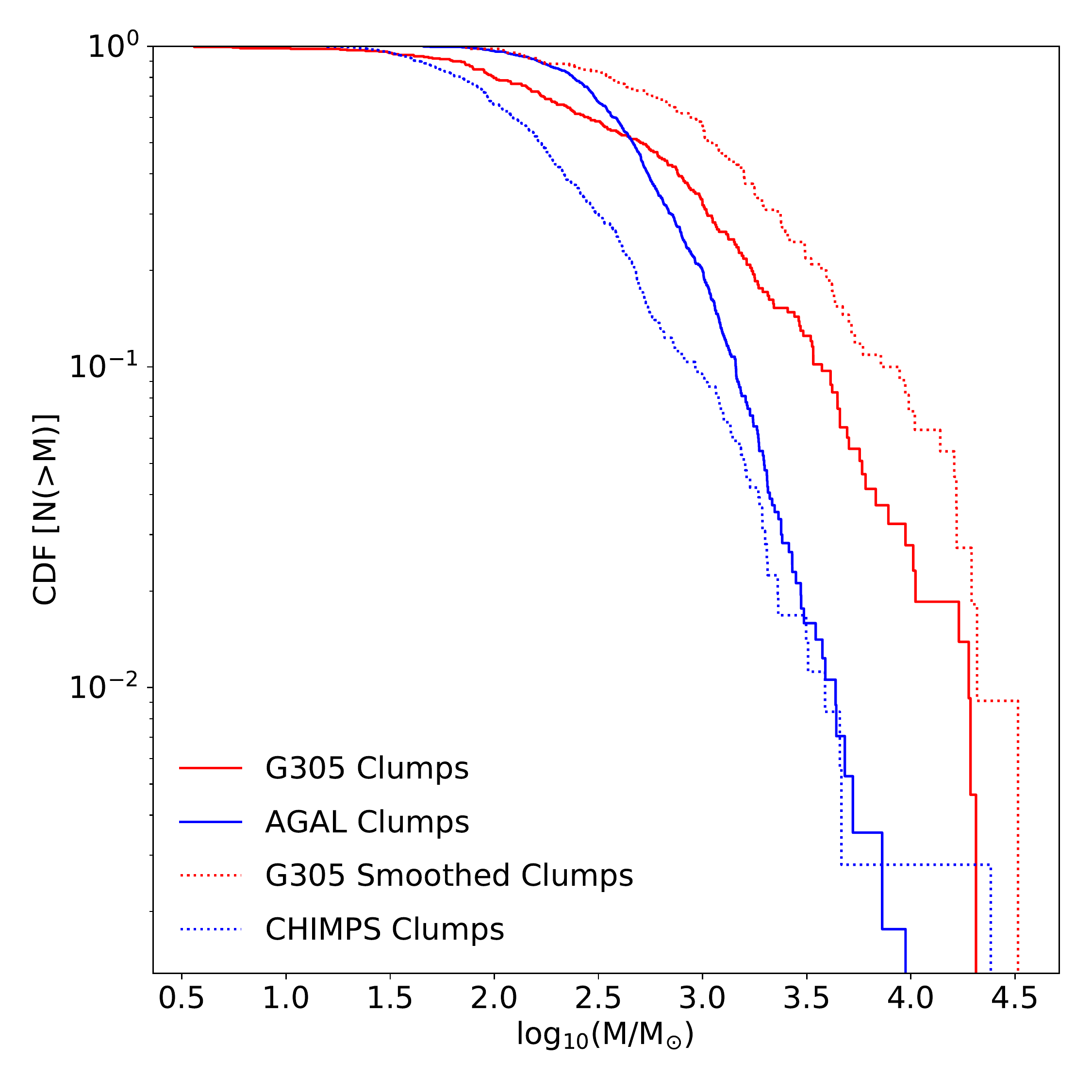}
            \caption{\texttt{[Sec.4]:} CDF of LAsMA clump masses in G305 compared to that of the Galactic clump sample from ATLASGAL and CHIMPS. \label{fig:mass_cdf}}
        \end{figure}

        \par Finally, in case of triggering, if C\&C is the dominant mechanism in play then we would observe more massive clumps as C\&C causes the clump to accumulate mass \citep{whitworth1994, deharveng2005}. Fig. \ref{fig:mass_cdf} shows the CDF of the clump masses in G305 compared to the Galactic average. In addition to the ATLASGAL, we also included the CHIMPS clumps. The method used to measure clump properties in our paper is identical to that used by \citet{rigby19}. Fig. \ref{fig:mass_cdf} shows the CDF of clump masses in G305 compared to the Galactic sample taken from ATLASGAL as well as CHIMPS. For both cases the clumps in G305 show a much flatter CDF than the Galactic average. A KS test between G305 clumps and ATLASGAL (CHIMPS) clumps yielded a p-value$<<0.0013$ again rejecting the null hypothesis that the G305 LAsMA clumps and the Galactic ATLASGAL (CHIMPS) clumps belong to the same distribution. The flatter slope of the clump mass CDF of G305 proves that C\&C is the dominant mechanism inside the GMC.

\section{Summary \label{sec:conc}}
    We observed the G305 region with the LAsMA receiver on APEX telescope using the \coaj and \cobj lines in order to test the effect of feedback from the central cluster of stars on the clump properties.
    \par We decomposed the molecular cloud into clumps using the Dendrogram analysis on the \cob datacube. We followed up by creating a catalog of clump properties properties. We tested the effect of feedback on two main clump properties namely linewidths and surface mass densities. Using the 8\,\mum emission map of the region as a proxy to feedback strength, we examined the correlation of the surface mass densities of the clumps with the incident average 8\,\mum flux. We then mask for the ``feedback zone" based on an 8\,\mum flux threshold of 100\,MJy/sr. The extent of overlap of each clump with the ``feedback zone" was used to determine whether it was mostly inside, partly inside or outside the feedback zone. The properties of these three populations of clumps were separately studied. We followed this up with a comparison of the properties of clumps in G305 as a whole to a distance limited sample of Galactic clumps taken from ATLASGAL and CHIMPS surveys assuming that the Galactic sample of clumps are similar to a population of clumps that are quiescent. The cumulative distribution function (CDF) of the masses as well as the $L/M$ ratios of the clumps (the latter having been derived entirely from ATLASGAL data) in G305 was obtained and compared to that of the Galactic average. We summarize our main findings below :
    
    \begin{itemize}
        \item Our data does not possess enough velocity resolution to discern the effect of feedback on the linewidths of the clumps.
        \item The surface mass densities of the clumps in the region are positively correlated to the incident average 8\,\mum flux.
        \item The surface mass densities of the clumps mostly inside the feedback zone are the largest, followed by that of the clumps partly inside and outside the feedback zone respectively indicating that feedback is triggering their the star forming ability.
        \item Clumps inside the feedback zone show much higher level of fragmentation than those partly inside or outside it.
        \item The probability density function (PDF) of the properties of G305 clumps are significantly different than those of the clumps derived from a distance limited Galactic sample of clumps indicating clear evidences of triggering.
        \item The CDF of both the masses and $L/M$ ratios of the clumps in G305 are significantly flatter than that of the Galactic average. This is a strong evidence for triggering (more specifically the collect and collapse mechanism), not redistribution or dispersion, being the dominant mechanism in G305.
    \end{itemize}
    
    Therefore, we obtain multiple evidences demonstrating that the feedback from the central cluster of stars in G305 is triggering the collapse of the molecular cloud complex.
    
\begin{acknowledgements}
We thank the staff of the APEX telescope for their assistance in observations. This work acknowledges support by The Collaborative Research Council 956, sub-project A6, funded by the Deutsche Forschungsgemeinschaft (DFG). This research made use of astrodendro, a Python package to compute dendrograms of Astronomical data (http://www.dendrograms.org/). Data handling in this work has been done using the \texttt{Astropy} python package \citep{astropy:2013,astropy:2018}. Parts of this work are based on observations made with the Spitzer Space Telescope, which is operated by the Jet Propulsion Laboratory (JPL), California Institute of Technology under a contract with NASA. This publication also made use of data products from the Midcourse Space Experiment. Processing of the data was funded by the Ballistic Missile Defense Organization with additional support from NASA Office of Space Science. This research has also made use of the NASA/ IPAC Infrared Science Archive, which is operated by the JPL, under contract with NASA. 
\end{acknowledgements}

\bibliographystyle{aa} 
\bibliography{bibfile} 

\begin{appendix}


\section{Velocity vs 8\,\mum}
 The linewidths of leaves (highest hierarchical structures) shows no dependence on the incident 8um flux whereas the branches do show a positive correlation with the 8\,\mum flux. This could be the result of the size versus linewidth relationship of the clumps. On dividing the linewidths by the size of the structures \citep{heyer2009} the leaves still do not show any strong dependence of this property on the incident 8\,\mum flux. But the branches do have a weak positive dependence on the incident 8\,\mum flux. This indicates that at larger scales, feedback from the stars does inject turbulence into the GMC. Similar results were also seen in \citetalias{mazumdar2021} where the stacked spectra showed a positive dependence of their skewness and kurtosis on the incident 8um flux.
 
 \begin{figure}
      \centering
      \includegraphics[width=0.5\textwidth]{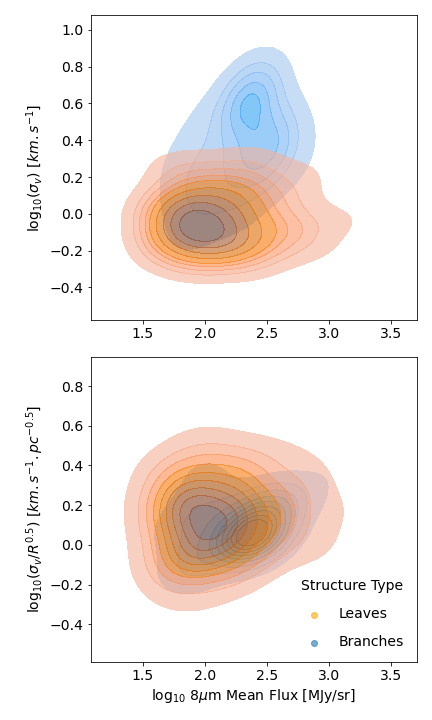}
      \caption{Linewidths for G305 clumps as a function of the incident 8\,\mum flux.}
      \label{fig:vrm_vs_8um}
  \end{figure}
\end{appendix}

\end{document}